\documentclass[3p,times]{elsarticle}

\usepackage{ecrc}

\volume{00}

\firstpage{1}

\journalname{Physics Procedia}

\runauth{}

\jid{phpro}

\jnltitlelogo{Physics Procedia}

\usepackage{amssymb}

\usepackage[figuresright]{rotating}

\begin{document}

\begin{frontmatter}



\dochead{}

\title{{\small12$^{\rm th}$ International Conference on Muon Spin Rotation, Relaxation and Resonance}\\
Magnetic order and transitions in the spin-web compound Cu$_3$TeO$_6$}


\author[ETH,PSI]{Martin~M\aa{}nsson\corref{cor1}},
\cortext[cor1]{Corresponding author. Tel.: +41-(0)56-310-5534 ; fax: +41-(0)44-633-1282}
\ead{mansson@phys.ethz.ch}
\author[ETH,PSI]{Krunoslav~Pr$\check{\rm s}$a}
\author[TCRDL]{Jun~Sugiyama}
\author[BBU]{Daniel~Andreica}
\author[LMU]{Hubertus~Luetkens}
\author[EPFL]{and Helmuth~Berger}

\address[ETH]{Laboratory for Solid state physics, ETH Z\"{u}rich, CH-8093 Z\"{u}rich, Switzerland}
\address[PSI]{Laboratory for Neutron Scattering, Paul Scherrer Institute, CH-5232 Villigen PSI, Switzerland}
\address[TCRDL]{Toyota Central Research and Development Labs. Inc., Nagakute, Aichi 480-1192, Japan}
\address[BBU]{Faculty of Physics, Babes-Bolyai University, 3400 Cluj-Napoca, Romania}
\address[LMU]{Laboratory for Muon-Spin Spectroscopy, Paul Scherrer Institut, CH-5232 Villigen PSI, Switzerland}
\address[EPFL]{Institute de Physique de la Matiere Complexe, EPFL, CH-1015 Lausanne, Switzerland}

\begin{abstract}
The spin-web compound Cu$_3$TeO$_6$, belongs to an intriguing group of materials where magnetism is governed by 3$d^{9}$ copper Cu$^{2+}$ ions. This compound has been sparsely experimentally studied and we here present the first investigation of its local magnetic properties using muon-spin relaxation/rotation ($\mu^{+}$SR). Our results show a clear long-range 3D magnetic order below $T_{\rm N}$ as indicated by clear zero-field (ZF) muon-precessions. At $T_{N}$~=~61.7~K a very sharp transition is observed in the weak transverse-field (wTF) as well as ZF data. Contrary to suggestions by susceptibility measurements and inelastic neutron scattering, we find no evidence for either static or dynamic (on the time-scale of $\mu^+$SR) spin-correlations above $T_{\rm N}$.
\end{abstract}


\end{frontmatter}


\section{Introduction}

Although magnetism is a quantum phenomenon, properties of most of the magnetic insulators can be understood using a semi-classical approach, i.e. the linear spin-wave theory. Especially in three-dimensional (3D) systems quantum effects are expected to be small and it may be asked whether they are at all relevant. The insulating tricopper-tellurate Cu$_{3}$TeO$_{6}$ belongs to an intriguing group of compounds where the magnetism is governed by 3$d^{9}$ copper Cu$^{2+}$ ions \cite{Book}. This compound can be indexed by the cubic space group $Ia\overline{3}$ [$a$~=~9.537(1)~{\AA}] \cite{Hostachy, Falck}. As shown in Fig.~1, the unit cell consists of 8 regular TeO$_{6}$ octahedra and 24 copper ions \cite{Falck}. Each copper ion is surrounded by six oxygen ions forming a very irregular octahedron. There have been very few experimental studies of Cu$_{3}$TeO$_{6}$ and in fact only one published investigation is to be found regarding its magnetic properties \cite{Herak}. The susceptibility-vs.-T [$\chi(T)$] curve [inset of Fig.~2(b)] displays a kink at $T\approx60$~K indicating long-range magnetic ordering. From a wider temperature range, a good fit of $\chi^{-1}$ to a Curie-Weiss law {\bf (CWL)} for $T$~=~180~-~330~K gives $\Theta{}_{CW}$~=~-134~K [Fig.~2(b)]. This shows that the Cu$^{2+}$ spins are in fact strongly antiferromagnetically (AF) coupled. It is further evident that $\chi(T)$ also starts to deviate from CWL below approximately 180~K, possibly indicating the onset of a short-range spin order. To display a deviation from CWL as high as 3$T_{\rm N}$ is rather unusual for a 3D system and is only expected to be observed in low-dimensional and frustrated spin systems. The explanation is found in the magnetic structure that consists of a 3D ``spin-web'' lattice of $S=1/2$ Cu$^{2+}$ ions characterized by low connectivity. Here each spin has only 4 nearest neighbors and 4 next-nearest neighbors, which is comparable to the two-dimensional square lattice antiferromagnet (2DSL-AF). Further, the magnetization vs. magnetic field ($H$) curve displays anomalous behavior that cannot be explained by a simple spin-flip or spin-flop transition, but rather by field induced rotation of magnetic domains. Such situation is also supported by torque measurements \cite{Herak}.
\begin{figure}
\begin{center}
\includegraphics[keepaspectratio=true,width=120 mm]{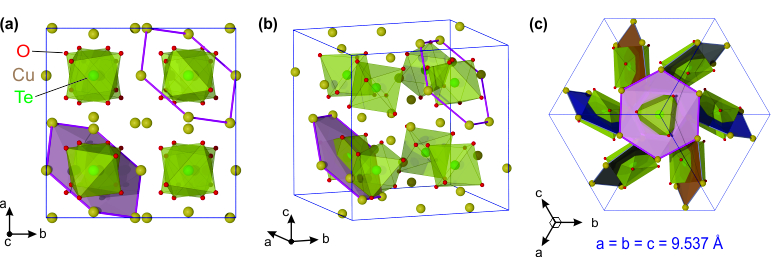}
\caption{\label{fig:1}The crystal structure of Cu$_{3}$TeO$_{6}$ [cubic space group $Ia\overline{3}$, $a$~=~9.537(1)~{\AA}] viewed along different directions. (c) Emphasizes the four different domains created by the copper planes oriented perpendicular to the space diagonals.}
  \label{fig:structure}
\end{center}
\end{figure}
Neutron powder diffraction show clear magnetic peaks below $T_{\rm N}$ \cite{Herak} and the magnetic structure is found to be a collinear (possibly canted) N\'{e}el order on a bipartite lattice with a wave vector $\overrightarrow{k}$~=~(0,~0,~0). The ordered moment determined by neutron diffraction is predominantly directed along the ($\pm1, \pm1, \pm1$) space diagonals and is found to be only 0.64 $\mu_{B}$/ion, i.e. strongly reduced from the classical 1 $\mu_{B}$/ion, again similar to the $S=1/2$ 2DSL-AF. Further, a recent inelastic neutron scattering (INS) study of the magnon dispersion in Cu$_{3}$TeO$_{6}$ show a large amount of diffuse scattering for $T~>~T_{N}$ \cite{Zaharko}. In similarity to the deviation from CWL this could possibly indicate the presence of short-range spin correlations above the bulk magnetic transition. Muon-spin relaxation/rotation ($\mu^{+}$SR) is an optimal technique to study such microscopic magnetic properties and we here present, to the best of our knowledge, the first $\mu^{+}$SR investigation of the Cu$_{3}$TeO$_{6}$ compound.

\begin{figure}
\begin{center}
\includegraphics[keepaspectratio=true,width=110 mm]{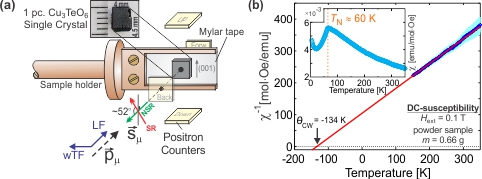}
\caption{\label{fig:1}(a) Schematic view of the experimental setup at the GPS spectrometer showing the direction of magnetic fields for spin-rotated (SR) and non-spin-rotated (NSR) mode. (b) Temperature dependence of the magnetic susceptibility, $\chi$ (inset) and $\chi^{-1}$. A clear kink at $T_{\rm N}\approx60$~K indicates the onset of long-range magnetic ordering. Solid red line is a fit to Curie-Weiss law yielding $\Theta{}_{CW}$~=~-134~K.}
\end{center}
\end{figure}

\section{\label{sec:E}Experimental Details}
Single crystals of Cu$_{3}$TeO$_{3}$ can be obtained by a HBr chemical transport method resulting in prisms of approximately 2$\times$2 mm, or even bigger. For this $\mu^+$SR experiment, a single piece of single crystal (4$\times$4$\times$2~mm$^{3}$) was attached to a low-background (fork-type) sample holder [see Fig.~2(a)] using very thin Al-coated Mylar tape. In order to make certain that the muon stopped primarily inside the sample, we ensured that the side facing the muon beamline was only covered by a single layer of mylar tape. Subsequently, $\mu^+$SR spectra were measured at the Swiss Muon Source (S$\mu$S), Paul Scherrer Institut, Villigen, Switzerland. By using the surface muon beam-line $\pi$M3.2, weak transverse-field (wTF) and zero-field (ZF) spectra were collected (see Fig.~3 \& 4) at the General Purpose Spectrometer (GPS). Data was collected for both non-spin-rotated mode (NSR) and spin-rotated (SR) mode [see Fig.~2(a)]. The experimental setup and techniques are described in greater detail elsewhere \cite{Kalvius}.

\section{\label{sec:E}Results \& Discussion}
\begin{figure}
\begin{center}
\includegraphics[keepaspectratio=true,width=145 mm]{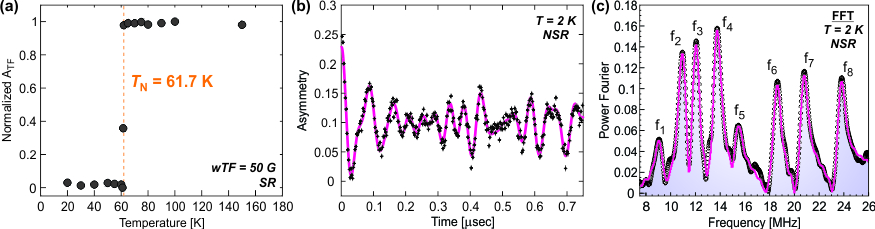}
\caption{\label{fig:1}(a) Temperature dependence of the wTF asymmetry ($A_{\rm TF}$) showing a sharp transition at $T_{\rm N}=61.7$~K. (b) ZF $\mu^{+}$SR time spectrum at $T=2$~K using NSR mode. Fast Fourier transform (FFT) of (b) showing eight different frequencies ($f_{i}$). Solid lines in (b-c) are a fits to Eq.(1).}
\end{center}
\end{figure}
In order to obtain an overview of the magnetic transition weak transverse-field (wTF = 50 G) $\mu^{+}$SR spectra were obtained as a function of temperature. In Fig.~3(a) the $T$-dependence of the wTF precession asymmetry ($A_{\rm TF}$) is shown where $A_{\rm TF}$ is completely suppressed below $T_{\rm N}$. At $T_{\rm N}$~=~61.7~K a clear bulk magnetic transition occurs, in accordance with susceptibility measurements, and $A_{\rm TF}$ quickly reaches its expected maximum value. That the full $A_{\rm TF}$ is recovered immediately above $T_{\rm N}$ is a strong indication that no magnetic order exists above $T_{\rm N}$. However, to gain a more robust verification also zero-field (ZF) $\mu^{+}$SR data was acquired. Below $T_{\rm N}$ a clear static magnetic order is seen from the obvious oscillation in the ZF time spectrum [Fig.~3(b)] acquired at $T=2$~K. From a fast Fourier transform (FFT) of the time spectrum [Fig.~3(c)], eight distinct muon precession frequencies ($f_{i}$, where $i=1..8$) can be decerned. Consequently, the ZF data was well fitted by the combination of eight damped cosine oscillations and a \textit{tail} signal due to the field component parallel to the initial muon spin:
\begin{eqnarray}
 A_0 \, P_{\rm ZF}(t) = A_{\rm tail}~e^{-\lambda_{\rm tail} t} + \sum^{8}_{i=1}A_{\rm i}~e^{-\lambda_{\rm i} t}\cos(2\pi f_{\rm i}\cdot{}t+\phi_i)~,
\label{eq:ZFfit}
\end{eqnarray}
In order to further investigate the formation of static magnetic order ZF-$\mu^+$SR spectra were collected for 2~K~$\leq{}T\leq{}$~70~K. As seen from Fig.~4(a) and from the temperature dependencies of $f_{i}$ [Fig.~4(c)], as $T$ increases the oscillation frequencies are gradually slowing down and finally disappears around $T_{\rm N}$. The magnetic order parameter [Fig.~4(c)] was found to be well fitted to the commonly used phenomenological formula:
\begin{eqnarray}
 f_{i}(T) = f_{\rm T\rightarrow 0 K} \cdot \left[1-\left(\frac{T}{T_{\rm N}}\right)^{\alpha}\right]^{~\beta},
\label{eq:ZFfit}
\end{eqnarray}
where $\beta$ is a parameter describing the dimensionality of the magnetic order \cite{Collins}. In these fits $T_{\rm N}$ was fixed to 61.7~K as obtained from the wTF measurements and the values of $\beta$ was shown to range from 0.4-0.5. Contrary to the proposed 2DSL-AF nature, on the length-scale of $\mu^+$SR, such $\beta$-values indicate a fully 3D (possibly Heisenberg) type magnetic order \cite{Collins}. Further, that all $f_{i}$'s show the same temperature evolution (i.e. similar $\beta$ and transition temperature) suggests that the multiple frequencies are not caused by the coexistence of different magnetic phases but rather a result of several inequivalent interatomic muon stopping sites within the crystallographic lattice.
\begin{figure}
\begin{center}
\includegraphics[keepaspectratio=true,width=145 mm]{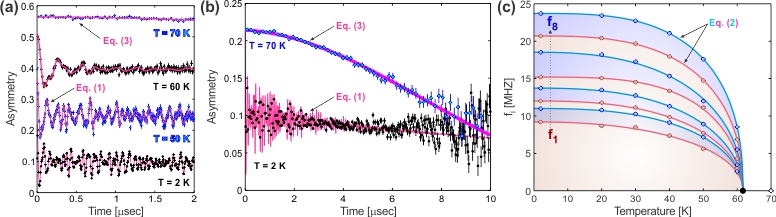}
\caption{(a-b) Temperature dependence of ZF time spectra using SR mode. Solid lines are fits to Eq.~(1) and Eq.~(3). (b) Magnetic order parameter showing the temperature dependence of the oscillation frequencies ($f_{i}$ where $i=1..8$). Solid lines are fits to Eq.~(2).}
\end{center}
\end{figure}
Noteworthy is also that the initial phases [$\phi_{i}$ in Eq.~(1)] of the precession frequencies show (if not fixed) strong deviations from zero and in addition complex $T-$dependencies. Such non-zero phases along with a rather asymmetric field distribution [see e.g. the FFT spectrum in Fig.~3(c)] could be an indication for incommensurate magnetic order in this compound. However, it is more likely that such phase-delays are caused by a wide field distribution due to the large amount of inequivalent muon sites, as we have previously reported for the LiCrO$_{2}$ compound \cite{LiCrO2}. This is clearly also in line with the neutron diffraction data that indeed show a clear commensurate order \cite{Herak}.

Finally, the ZF data obtained just above $T_{\rm N}$ [$T$~=~70~K in Fig.~4(a-b)] show no indication of short-range magnetic order. Instead it is well fitted by a simple static Gaussian Kubo-Toyabe (KT) function [$G^{\rm KT}(\Delta_{\rm KT},t)$] multiplied by an exponential relaxation:
\begin{eqnarray}
 A_0 \, P_{\rm ZF}(t) = A_{\rm KT}~G^{\rm KT}(\Delta_{\rm KT},t)\cdot e^{-\lambda_{\rm elec} t}~,
\label{eq:KT}
\end{eqnarray}
Here $G^{\rm KT}(\Delta_{\rm KT},t)$ originates from the presence of randomized (nuclear) magnetic moments. From such fit we obtain $\Delta_{\rm KT}=0.095$~MHz along with $\lambda_{\rm elec}=0.009$~MHz for $T=70$~K. The negligible value of $\lambda_{\rm elec}$ indicates the absence of any strong fluctuations in the electronic moments (spins).

\section{\label{sec:E}Summary}
We here present the first study of local magnetic properties of the spin-web compound Cu$_3$TeO$_6$. On the length-scale of $\mu^+$SR, both wTF and ZF measurements show a clear long-range 3D magnetic order below $T_{N}$~=~61.7~K. Contrary to suggestions by susceptibility measurements and inelastic neutron scattering, we further find no clear evidence for either static or dynamic (on the time-scale of $\mu^+$SR) spin-correlations above $T_{\rm N}$.

\paragraph{\textbf{Acknowledgments}}\

This work was performed using the \textbf{GPS} muon spectrometer at the Swiss Muon Source (S$\mu$S) of the Paul Scherrer Institut (PSI), Villigen, Switzerland and we are thankful to the instrument staff for their support. All the $\mu^{+}$SR data was fitted using \texttt{musrfit} \cite{musrfit} and the images involving crystal structure were made using the DIAMOND software. This research was financially supported by the Swiss National Science Foundation (through Project 6, NCCR MaNEP) and Toyota Central Research \& Development Labs. Inc. DA acknowledges financial support from the Romanian CNCSIS-UEFISCU Project PNIIIDEI 2597/2009 (Contract No. 444).

\paragraph{\textbf{References}}\





\bibliographystyle{elsarticle-num}

\begin{thebibliography}{1}
\expandafter\ifx\csname url\endcsname\relax
  \def\url#1{\texttt{#1}}\fi
\expandafter\ifx\csname urlprefix\endcsname\relax\def\urlprefix{URL }\fi
\expandafter\ifx\csname href\endcsname\relax
  \def\href#1#2{#2} \def\path#1{#1}\fi

\bibitem{Book}
Quantum Magnetism (Springer Lecture Notes in Physics), Berlin: Springer, 2004.

\bibitem{Hostachy}
{A. Hostachy and J. Coing-Boyat}, C. R. Acad. Sci. 267 (1968) 1435.

\bibitem{Falck}
{L. Falck, O. Lindqvist and J. Moret}, Acta Crystallogr. B 34 (1978) 896.

\bibitem{Herak}
{M. Herak $et~al.$}, J. Phys.: Condens. Matter 17 (2005) 7667.

\bibitem{Zaharko}
{O. Zaharko $et~al.$}, $private~communication$ (2011).

\bibitem{Kalvius}
{G. M. Kalvius, D. R. Noakes, and O. Hartmann}, Handbook on the Physics and
  Chemistry of Rare Earths, Vol.~32, Elsevier Science B. V. Amsterdam, 2001,
  Ch. 206.

\bibitem{Collins}
{M. F. Collins}, Magnetic critical scattering, Oxford Univ. Press, Oxford,
  1989, Ch. 5, table 5.1.

\bibitem{LiCrO2}
{J. Sugiyama, M. M\aa nsson, Y. Ikedo $et~al.$}, Phys. Rev. B 79 (2009) 184411.

\bibitem{musrfit}
{A. Suter and B. M. Wojek}, $in~these~proceedings$ (2011).

\end{thebibliography}



\end{document}